\documentclass{elsart}

\usepackage{amsmath,amssymb,graphicx,bm}

\newcommand{\beqn}{\begin{equation}}
\newcommand{\eeqn}{\end{equation}}
\newcommand{\bea}{\begin{eqnarray}}
\newcommand{\eea}{\end{eqnarray}}

\newcommand{\vlowk}{V_{{\rm low}\,k}}
\newcommand{\vsrg}{V_s}

\newcommand{\fmi}{\, \text{fm}^{-1}}
\newcommand{\mev}{\, \text{MeV}}

\newcommand{\openone}{\leavevmode\hbox{\small1\normalsize\kern-.33em1}}

\newcommand{\qvec}{{\bf q}}

\newcommand{\kmax}{k_{\rm max}}

\newcommand{\adaggera}{a^\dagger_{\qvec} a^{{\phantom{\dagger}}}_{\qvec}}
\newcommand{\Hzero}{T_{\rm rel}}
\newcommand{\flow}{s}

\begin{document}

\begin{frontmatter}

\title{Are low-energy nuclear observables\\
sensitive to high-energy phase shifts?}

\author{S.K.\ Bogner}$^1$,
\ead{\\bogner@mps.ohio-state.edu}
\author{R.J.\ Furnstahl}$^1$,
\ead{\\furnstahl.1@osu.edu}
\author{R.J.\ Perry}$^1$,
\ead{\\perry@mps.ohio-state.edu}
\author{A.\ Schwenk}$^{2}$
\ead{\\schwenk@triumf.ca}
\address{$^1$Department of Physics,
The Ohio State University, Columbus, OH\ 43210
$^2$TRIUMF, 4004 Wesbrook Mall, Vancouver, BC, Canada, V6T 2A3}


\begin{abstract}
Conventional nucleon-nucleon potentials with strong short-range
repulsion require contributions from high-momentum wave function components 
even for low-energy observables such as the deuteron binding energy.
This can lead to the misconception that reproducing high-energy phase
shifts is important for such observables.
Interactions derived via the similarity renormalization group 
decouple high-energy and low-energy physics while preserving the
phase shifts from the starting potential.
They are used to show that high-momentum components (and high-energy
phase shifts) can be set to zero when using low-momentum
interactions, without losing information relevant 
for low-energy observables.
\end{abstract}

\end{frontmatter}
\maketitle

\section{Introduction}
\label{sec:intro}

High-momentum  degrees of freedom do not automatically decouple from
low-energy  observables, especially for conventional nucleon-nucleon
(NN) potentials. 
For instance, nuclear forces are typically fit to scattering 
data up to where inelasticities start to become significant, $E_{\rm lab} 
\sim 350 \mev$ or relative momenta $k \sim 2 \fmi$. However, most NN 
potentials have significant 
high-momentum ($k > 2 \fmi$) 
off-diagonal matrix elements that require summations over 
high-energy intermediate states, even if one is calculating low-energy 
observables.
If such interactions are simply truncated at $2 \fmi$, the deuteron binding 
energy 
along with S-wave phase shifts down to zero energy are drastically altered.  
This can lead to the misconception that \emph{details} of
strong-interaction dynamics above some energy scale are relevant to
low-energy nuclear structure and reactions.
Such a brute-force  cutoff, however,  does not disentangle high-energy
(short-distance) features from low-energy (long-distance)
observables. 
To do so, it is necessary to integrate out (and thus separate) 
irrelevant short-distance details from their effects on low-energy 
observables. This is achieved by the renormalization group.

Renormalization group (RG) transformations that lower a cutoff 
in relative momentum have been used to derive NN potentials
that have vanishing matrix elements for momenta above the cutoff.
Such interactions, known generically as $\vlowk$,
show greatly enhanced convergence properties in nuclear few- and 
many-body systems for cutoffs of order $\Lambda = 2\fmi$ or
lower~\cite{Vlowk1,Vlowk2,VlowkRG,Vlowk3N,Bogner_nucmatt,VlowkSM1,%
VlowkSM2}.
However, the initial NN potentials (typically cut off at $4$--$5 \fmi$)
predict non-zero elastic phase shifts to much higher energies
than $E_{\rm lab} \sim 350 \mev$, 
which in some channels are semi-quantitatively consistent with 
experiment. In contrast, phase shifts from $\vlowk$ as usually 
implemented are zero above the cutoff. 
This discrepancy has led to some uneasiness 
that important information may be lost with $\vlowk$.
Similar concerns have been expressed about effective field theory 
(EFT) potentials.

This unease is exacerbated by experience with conventional NN
potentials, which feature strong short-range repulsion.
The repulsion causes even bound states with very low energies 
(such as the deuteron) to have important contributions to the binding
and other properties from high-momentum components (well above $2\fmi$)
of the deuteron wave function.
In Ref.~\cite{Benhar:1993ja}, the authors calculate cross sections
for electron scattering from the deuteron using as input a spectral function
that is the momentum distribution times a delta function in
energy.
After noting the effect of excluding high momenta
on the cross section, they conclude:
``and thus the data confirm the existence of high-momentum components
in the deuteron wave function'' \cite{Benhar:1993ja}. 
Beyond the fact that wave functions are not observables, these conclusions 
reinforce the intuition that there is information in
quantitatively reproducing high-energy phase shifts that is lost
when evolving to low-momentum interactions.
In this paper, 
we use the similarity renormalization group (SRG) 
\cite{Glazek:1993rc,Wegner:1994,Szpigel:2000xj,Bogner:2006srg} as a tool to 
demonstrate unequivocally that this intuition is incorrect.

A fundamental tenet of renormalization theory is that the 
\emph{relevant} details of high-energy physics for calculating
low-energy observables can be captured in the scale-dependent coefficients
of operators in a low-energy hamiltonian \cite{Lepage}.
This principle does not mean that high-energy and low-energy physics
is entirely decoupled in an effective theory.
In fact, it implies that  we can include as much
irrelevant coupling to \emph{incorrect} high-energy physics as we want by using
a large cutoff, with no consequence
to low-energy predictions (assuming we can calculate accurately).
But this freedom also offers the possibility of decoupling, 
which makes practical
calculations more tractable by restricting the necessary degrees of freedom.
This decoupling can be efficiently achieved by evolving nuclear
interactions using RG transformations.
(For an earlier discussion of decoupling based on Okubo unitary
transformations, see Ref.~\cite{Epelbaum:1999}.)

The SRG allows for particularly transparent
and convincing demonstrations of
decoupling, because the evolution of the hamiltonian and
other operators proceeds via transformations that can be chosen
to be unitary,
so that all observables are explicitly preserved.
Thus, when we include the full set of momentum states used for the
original potential with the evolved potential $\vsrg$,
we find the same deuteron binding energy and phase shifts for 
all energies. However, the evolved $\vsrg$ explicitly 
decouples high-energy dynamics from low-energy observables, which 
means that we can exclude the high-momentum parts 
(so that we have a low-momentum potential like $\vlowk$) 
without disturbing the information content (phase shifts 
and deuteron binding energy) at lower energies. 
We apply a similar test to the SRG-evolved deuteron
momentum distribution to show that high-momentum effects in 
low-energy bound states are captured by scale-dependent 
low-momentum operators.
 
\section{Background on the SRG}
\label{sec:operators}

In Ref.~\cite{Bogner:2006srg}, the similarity renormalization group
(SRG) approach is applied to NN interactions. The SRG was developed 
by Glazek and Wilson~\cite{Glazek:1993rc}, while Wegner~\cite{Wegner:1994} 
independently developed a related but simpler set of flow equations.
The formalism we employ closely resembles that of Wegner, but we find 
that a transformation even simpler than advocated by 
Wegner is robust and adequate for all calculations to date.

The evolution or flow of the 
hamiltonian with a parameter $s$ follows from a unitary transformation, 
\beqn
H_\flow = U(\flow) H U^\dagger(\flow) \equiv \Hzero + V_\flow \,,
\label{eq:Hflow}
\eeqn
where $\Hzero$ is the
relative kinetic energy and $H = \Hzero + V$ is the initial 
hamiltonian in the center-of-mass system. 
Eq.~(\ref{eq:Hflow}) defines the evolved
potential $V_\flow$, with $\Hzero$ taken to be independent of $\flow$.
Then $H_\flow$ evolves according to
\beqn
\frac{dH_\flow}{d\flow} = [\eta(\flow),H_\flow] \,,
\eeqn
with
\beqn
\eta(\flow) = \frac{dU(\flow)}{d\flow} U^\dagger(\flow) 
= -\eta^\dagger(\flow) \,.
\eeqn
Choosing $\eta(\flow)$ specifies the transformation.
As in Ref.~\cite{Bogner:2006srg}, 
we restrict ourselves to the simple choice~\cite{Szpigel:2000xj}
\beqn
\eta(\flow) =  [\Hzero, H_\flow] \,,
\label{eq:choice}
\eeqn
which gives the flow equation,
\beqn
\frac{dH_\flow}{d\flow} = [ [\Hzero, H_\flow], H_\flow] 
= [ [\Hzero, V_\flow], H_\flow] \,.
\label{eq:commutator}
\eeqn 
In a momentum basis,
this choice suppresses off-diagonal matrix elements, forcing the
hamiltonian towards a band-diagonal form.

The evolution in Eq.~(\ref{eq:commutator}) includes all many-body
components of the hamiltonian.
In the space of relative momentum NN states only, 
it means that the partial-wave momentum-space potential evolves
as (with normalization so that $1 = \frac{2}{\pi} \int_0^\infty\! q^2 \, 
dq \, |q\rangle \langle q|$ and in units where $\hbar = c = m = 1$
with nucleon mass $m$),
\bea
\frac{dV_\flow(k,k')}{d\flow} &=& - (k^2 - k'{}^2)^2 \, V_\flow(k,k')
\nonumber \\[2mm]
&& + \frac{2}{\pi} \int_0^\infty\! q^2 \, dq \, (k^2 + k'{}^2 - 2q^2)
\, V_\flow(k,q) \, V_\flow(q,k') \,.
\label{eq:diffeq}
\eea
(The additional matrix structure of $V_\flow$ in coupled channels
such as $^3$S$_1$--$^3$D$_1$ is implicit.)
For matrix elements far from the diagonal, the first term
on the right side of Eq.~(\ref{eq:diffeq}) dominates and 
exponentially suppresses
these elements as $\flow$ increases.
By discretizing the relative momentum space on a grid of gaussian integration
points, 
we obtain a simple (but nonlinear) system of first-order
coupled differential equations,
with the boundary condition that $V_\flow(k,k')$ at the initial
$\flow = 0$ is equal to the initial potential.
Since the SRG transformation is unitary,
the NN phase shifts and deuteron binding energy
calculated with $H_\flow$ are independent
of $\flow$ to within numerical precision.

The evolution with $\flow$ of any other operator $O$ is given by the same
unitary transformation, 
$O_\flow = U(\flow) O U^\dagger(\flow)$,
which means that $O_\flow$ evolves according to
\beqn
\frac{dO_\flow}{d\flow}
= [\eta(\flow),O_\flow] = [ [\Hzero,V_\flow], O_\flow] \,.
\label{eq:opflow}
\eeqn
Just as with the hamiltonian $H_\flow$, this evolution will induce many-body
operators even if the initial operator is purely two-body.
If we restrict ourselves to the relative momentum NN space, we have 
\bea
\frac{dO_\flow(k,k')}{d\flow} &=& \frac{2}{\pi} \int_0^\infty\! q^2 \, dq
\, \Bigl[ (k^2 - q^2) \, V_\flow(k,q) \, O_\flow(q,k')
\nonumber \\[2mm]
&& + (k'{}^2 - q^2) \, O_\flow(k,q)\, V_\flow(q,k') \Bigr] \,.
\label{eq:Odiffeq}
\eea
To evolve a particular $O_\flow$ simultaneously with $V_\flow$, 
we simply include
the discretized version of Eq.~(\ref{eq:Odiffeq}) as additional
coupled first-order differential equations.

An alternative, more direct, approach in the two-body space 
is to construct the unitary transformation at each $s$ explicitly, and then
use it to transform operators.
Let $| \psi_\alpha(s) \rangle$ be an eigenstate of $H_\flow$
with eigenvalue $E_\alpha$ (which is independent of $\flow$).
Then $U(s)$ is given by
\beqn
U(s) = \sum_\alpha | \psi_\alpha(s) \rangle \langle \psi_\alpha(0) | \,.
\label{eq:Ualpha}
\eeqn
In a discretized partial-wave relative-momentum space with momenta
$\{k_i\}$, we can solve for the eigenvectors
of $H = H_{s=0}$ and $H_s$, then construct the matrix elements of
$U(s)$ [which we denote $U_s(k_i,k_j)]$
by summing over the product of momentum-space wave functions:
\beqn
U_s(k_i,k_j) = \sum_\alpha \, \langle k_i | \psi_\alpha(s) \rangle  
\langle \psi_\alpha(0) | k_j \rangle \,.
\eeqn
In practice this is an efficient way to construct the
unitary transformation and subsequently to evolve any operator 
in the two-body space.

\section{Applying the SRG}
\label{sec:results}

In Ref.~\cite{Bogner:2006srg}, the SRG was applied with chiral EFT 
interactions at N$^3$LO~\cite{N3LO,N3LOEGM} as the initial potentials. 
It was shown that the simple 
SRG transformation drives the hamiltonian towards the diagonal
(in momentum space), making it more perturbative and more convergent in
few-body calculations. (In fact, it behaves just like a $\vlowk$
potential.) The parameter $\lambda \equiv s^{-1/4}$ provides a measure of
the spread of off-diagonal strength in $\vsrg$. The $\vsrg$ potential at
$\lambda$  was found to correspond roughly to a smoothly regulated
$\vlowk$ interaction~\cite{Vlowksmooth} 
with momentum cutoff $\Lambda \approx \lambda$. 
In this work, we also use the Argonne $v_{18}$~\cite{AV18} and the
CD-Bonn~\cite{CDBonn} potentials as initial potentials in the SRG 
evolution. 
We find the same universal behavior \cite{srgwebsite}.

\begin{figure}[tp]
\begin{center}
\includegraphics*[width=5.175in]{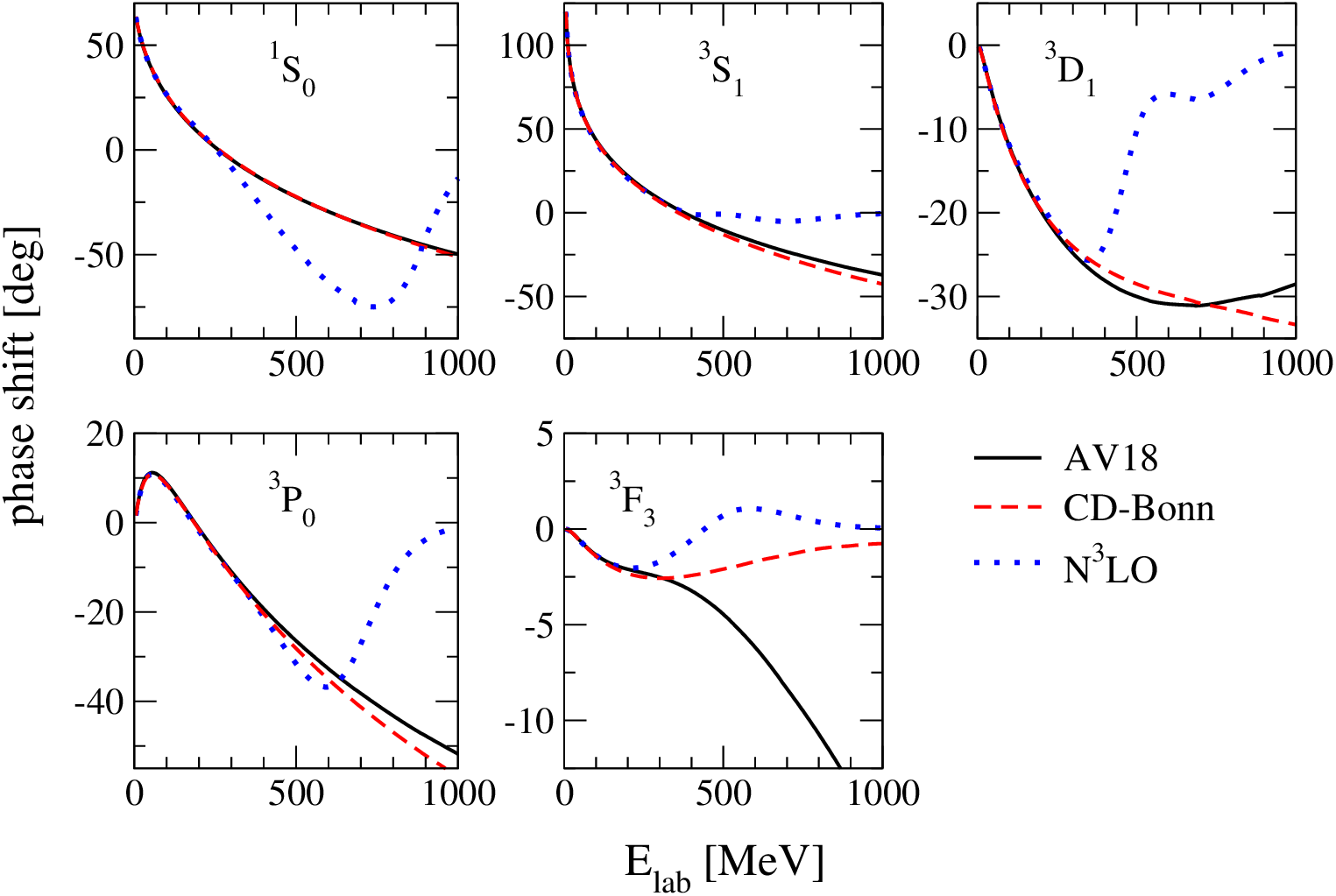}
\end{center}
\caption{Phase shifts for the Argonne $v_{18}$~\cite{AV18},
CD-Bonn~\cite{CDBonn} and one of the chiral N$^3$LO~\cite{N3LO} potentials
in selected channels (using nonrelativistic kinematics).
The phase shifts after evolving in $\lambda$ from each initial
potential agree for all $\lambda$ to within the widths of the
lines at all energies.}
\label{fig:phases}
\vspace*{.2in}
\begin{center}
\includegraphics*[width=5.36in]{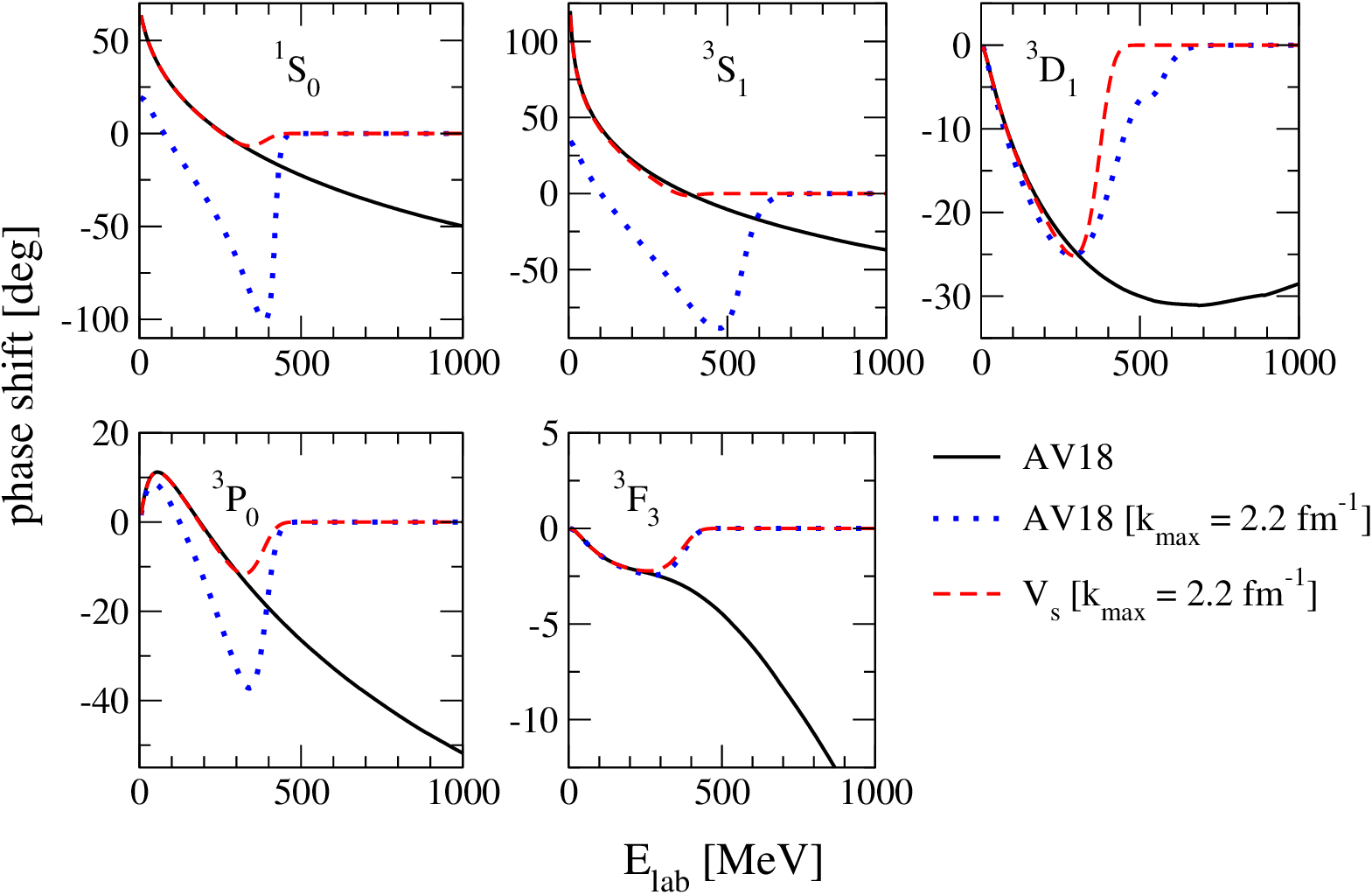}
\end{center}
\caption{Phase shifts in selected channels for
the Argonne $v_{18}$ potential~\cite{AV18} and when
intermediate momenta $k > \kmax = 2.2 \fmi$ are excluded. We
contrast the latter results to the phase shifts obtained from
the evolved $\vsrg$ potential for $\lambda = 2\fmi$ and the
additional constraint $\kmax = 2.2 \fmi$.}
\label{fig:phases_exclude}
\end{figure}

In Fig.~\ref{fig:phases}, phase shifts in selected partial waves
up to laboratory energies of $1 \, \text{GeV}$ are shown for the 
Argonne $v_{18}$~\cite{AV18}, CD-Bonn~\cite{CDBonn} and chiral 
N$^3$LO~\cite{N3LO} potentials.%
\footnote{Nonrelativistic kinematics is used, which means that comparisons
with experiment at higher energy should be made with caution. This can be 
improved but will not change our discussion because it
will have no effect on low-energy observables.}
As expected for unitary transformations, the phase shifts at any
$\lambda$ are the same as those for the 
corresponding initial potential (up to
numerical inaccuracies that are currently less than $0.1 \%$ in 
the worst case, with no real attempt at numerical optimization).
The phase shifts from different initial potentials
start to disagree for energies above $300$--$400 \mev$. The fact that
they lead to very similar low-momentum interactions~\cite{Vlowk1,%
Vlowk2,Bogner:2006srg} is an indication that all low-energy 
observables should similarly decouple after RG evolution. But here
we explicitly demonstrate this decoupling rather than relying on
generic RG arguments.

We test the decoupling of high-energy details from low-energy
phase shifts by setting $V_s(k,k')$ to zero for all $k,k'$ above
a specified momentum $\kmax$.
Results for $\kmax = 2.2 \fmi$ using a smooth regulator function 
are shown in
Fig.~\ref{fig:phases_exclude} for the initial Argonne $v_{18}$
potential and for the evolved $\vsrg$ potential with $\lambda 
= 2.0 \fmi$. The phase shifts for the initial potential 
in the lower partial waves bear 
no relation to the result without a $\kmax$ cutoff.
The 
coupling between high and
low momentum for the Argonne potential is so strong in   
the coupled $^3$S$_1$--$^3$D$_1$ channel
that we had to use a slightly larger
$\kmax = 2.5 \fmi$ with a smoother regulator 
just to keep the phase shifts on the plot.
In contrast, the low-energy
phase shifts for the SRG-evolved potential are unchanged,
even though the high-energy phase shifts above $\kmax$ are now zero. 
The details of how the momentum is cut off affect only the behavior
near $\kmax$.
The deviation near $\kmax$ in Fig.~\ref{fig:phases_exclude}
is consistent with the SRG (with the 
present choice of $\eta$) effectively imposing a rather smooth 
cutoff in momentum, similar to an exponential regulator with 
exponent $n_{\rm exp}=2$, which is the analytic behavior for solutions of 
the linearized SRG equation. 
Finally, the angular momentum barrier in higher partial waves such
as $^3$F$_3$ ensures that cutting off high 
momentum has the same effect for all interactions since they share the 
same long-range one-pion exchange potential.
 
\begin{figure}
\begin{center}
\includegraphics*[width=4.7in]{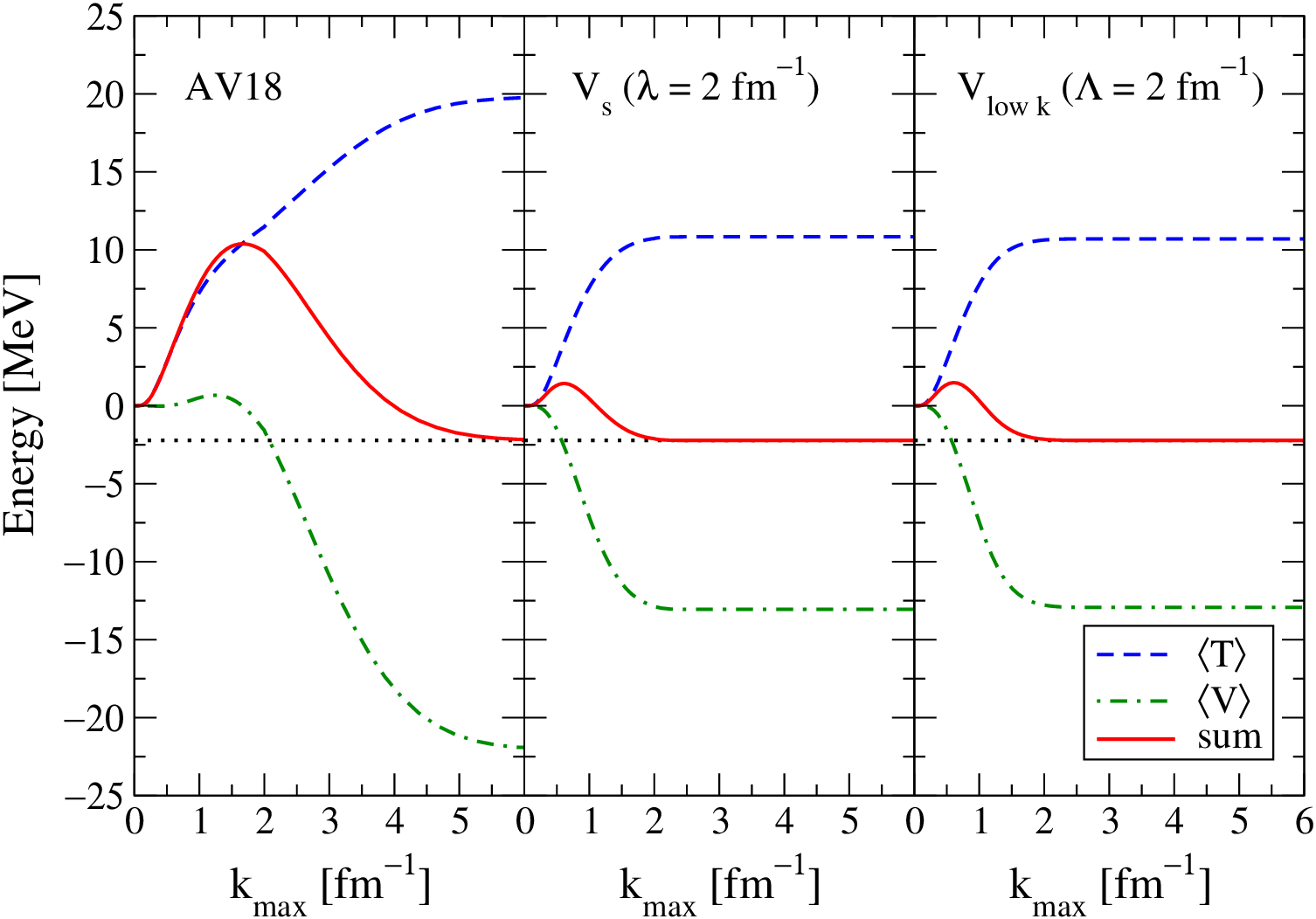}
\end{center}
\caption{Expectation values in the deuteron of the kinetic, 
potential, and total energy evaluated in momentum space as 
a function of the maximum momentum  $\kmax$, see
Eq.~(\ref{eq:Ed}). Results are shown
for the Argonne $v_{18}$ potential~\cite{AV18} (left), the 
evolved $\vsrg$ potential for $\lambda = 2\fmi$, and the
smooth-cutoff $\vlowk$ interaction with $\Lambda=2 \fmi$
and exponential regulator $n_{\rm exp}=2$}
\label{fig:deutbinding}
\end{figure}

The deuteron binding energy provides another clear example of 
how the contributions of different momentum components to a low-energy
observable depend on the resolution scale (as measured by $\lambda$
or $s$).
In Fig.~\ref{fig:deutbinding}, we show the  kinetic, potential, and 
total energy from an integration in momentum space including momenta 
up to $\kmax$. That is, we plot
\begin{align}
  E_d (k < \kmax) &= \Hzero(k < \kmax) + V_\flow(k < \kmax) \nonumber \\[2mm]
   &= \int_0^{\kmax}\! d{\bf k} \int_0^{\kmax}\! d{\bf k'}\,
  \psi_d^\dagger({\bf k};\lambda)\left( k^2 \delta^3({\bf k} - {\bf k'})
  + V_s({\bf k},{\bf k'}) \right)\psi_d({\bf k'};\lambda) \,,
  \label{eq:Ed}
\end{align}
where $\psi_d({\bf k};\lambda)$ is the momentum-space deuteron wave function
from the corresponding potential $V_s$ (without $\kmax$).

Figure~\ref{fig:deutbinding} shows that the Argonne $v_{18}$ 
potential~\cite{AV18} has large (and canceling) contributions
from the high-momentum matrix elements of the hamiltonian. 
For example,
if one excludes momenta greater than $2\fmi$ in the deuteron 
wave function when calculating the binding energy, 
the deuteron is $9.9 \mev$ unbound
(that is, the integrated kinetic energy up to $2\fmi$ is $11.5 \mev$ 
while the potential energy is $-1.6 \mev$). One needs to include 
contributions up to $4 \fmi$ to even get a bound state.
In contrast, using $\vsrg$ with $\lambda = 2\fmi$, we see that
the converged result is dominated by contributions from much 
lower momenta.%
\footnote{We can also simply set the high-momentum matrix elements of 
$\vsrg$ to zero (that is, exclude momenta larger than $\kmax$ 
at the level of the hamiltonian), and then solve for the deuteron.
For $\lambda = 2\fmi$, choosing $\kmax = 
2.2 \fmi$ only changes the binding energy to $2.2 \mev$, while 
raising $\kmax$ slightly yields the full $\vsrg$ result.}
Note that the $\vsrg$ potential has no appreciable contributions
above $\lambda$, even though the near-diagonal matrix 
elements of the potential $V_s(k,k')$ for $k,k' > \kmax$ are 
not negligible. This again validates the advertised decoupling.
We also see that the $\vsrg$ and $\vlowk$ results are very similar
for $\lambda \approx \Lambda$, where $\Lambda$ is the momentum cutoff
for $\vlowk$.

We turn to the momentum distribution in the deuteron next.
The momentum distribution at relative momentum ${\bf q}$ 
is the expectation value of $\adaggera$ 
(summed over spin substates $M_S$) and is proportional to the
sum of the squares of the normalized
S-wave and D-wave parts of the deuteron wave function,
$u(q)^2 + w(q)^2$.
It is not directly related to an observable (see, for example,
Ref.~\cite{Furnstahl:2001xq}).
As discussed in Sect.~\ref{sec:operators},
using the SRG we can consistently evolve operators
in $s$ (or $\lambda$).  In particular, we can evolve $\adaggera$
starting from 
a given hamiltonian.  
Since the SRG proceeds via
unitary transformations, no information is lost, by construction.
But even more,
we can show explicitly that by evolving to a low-momentum interaction,
we decouple the low-momentum and high-momentum physics in a low-energy
state.

\begin{figure}
\begin{center}
\includegraphics*[width=4.5in]{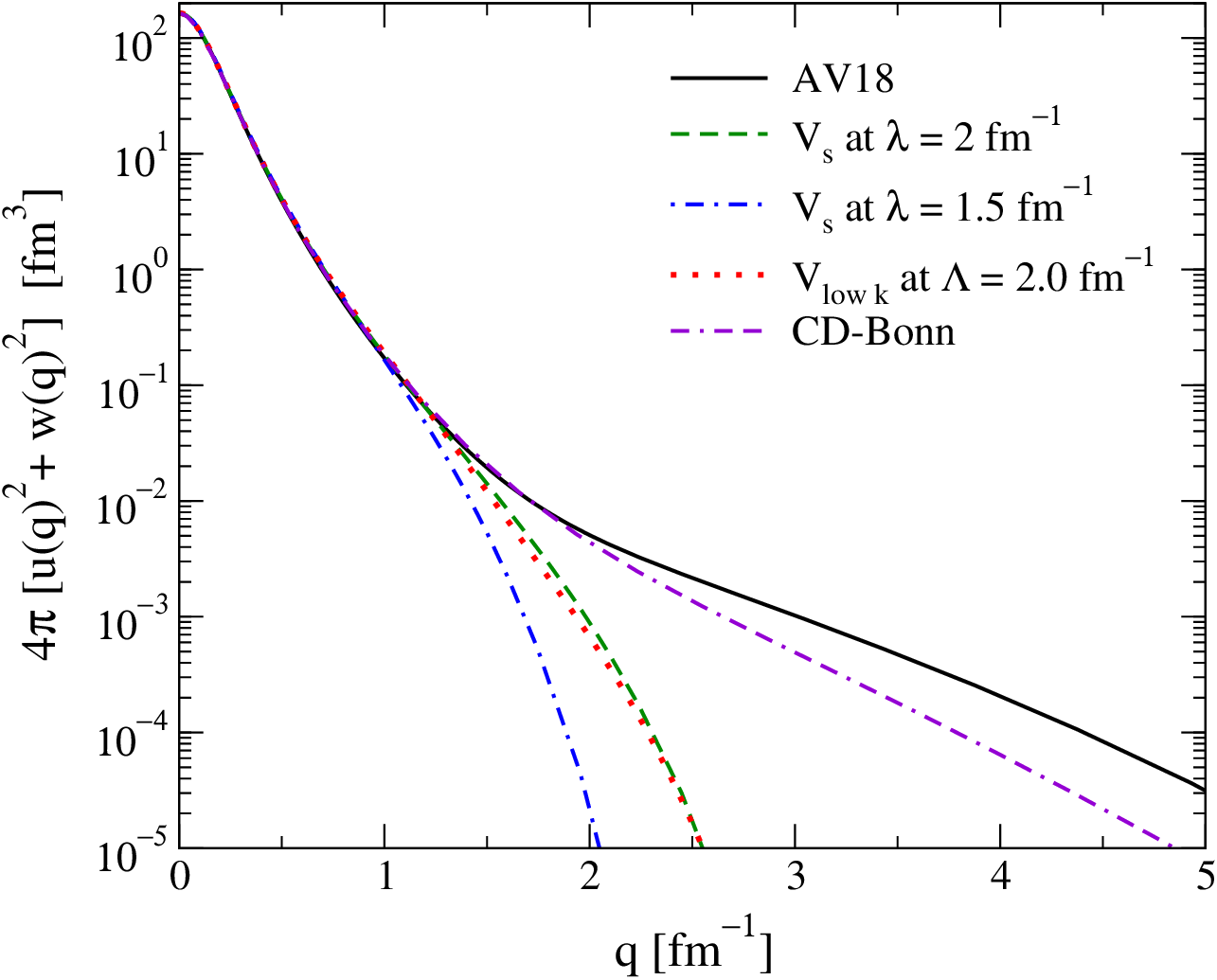}
\end{center}
\caption{Deuteron momentum distribution $\langle a^\dagger_{\qvec} 
a^{\protect\phantom\dagger}_{\qvec} \rangle_d \propto u(q)^2 + w(q)^2$
using the Argonne $v_{18}$~\cite{AV18}, CD-Bonn~\cite{CDBonn} and 
SRG potentials evolved from Argonne $v_{18}$ to $\lambda = 1.5\fmi$ 
and $2\fmi$ (but not evolving the operator), and a smooth-cutoff
$\vlowk$ interaction with $\Lambda = 2\fmi$ and exponential regulator
$n_{\rm exp}=2$.}
\label{fig:deutmd}
\end{figure}
 
\begin{figure}
\begin{center}
\includegraphics*[width=4.5in]{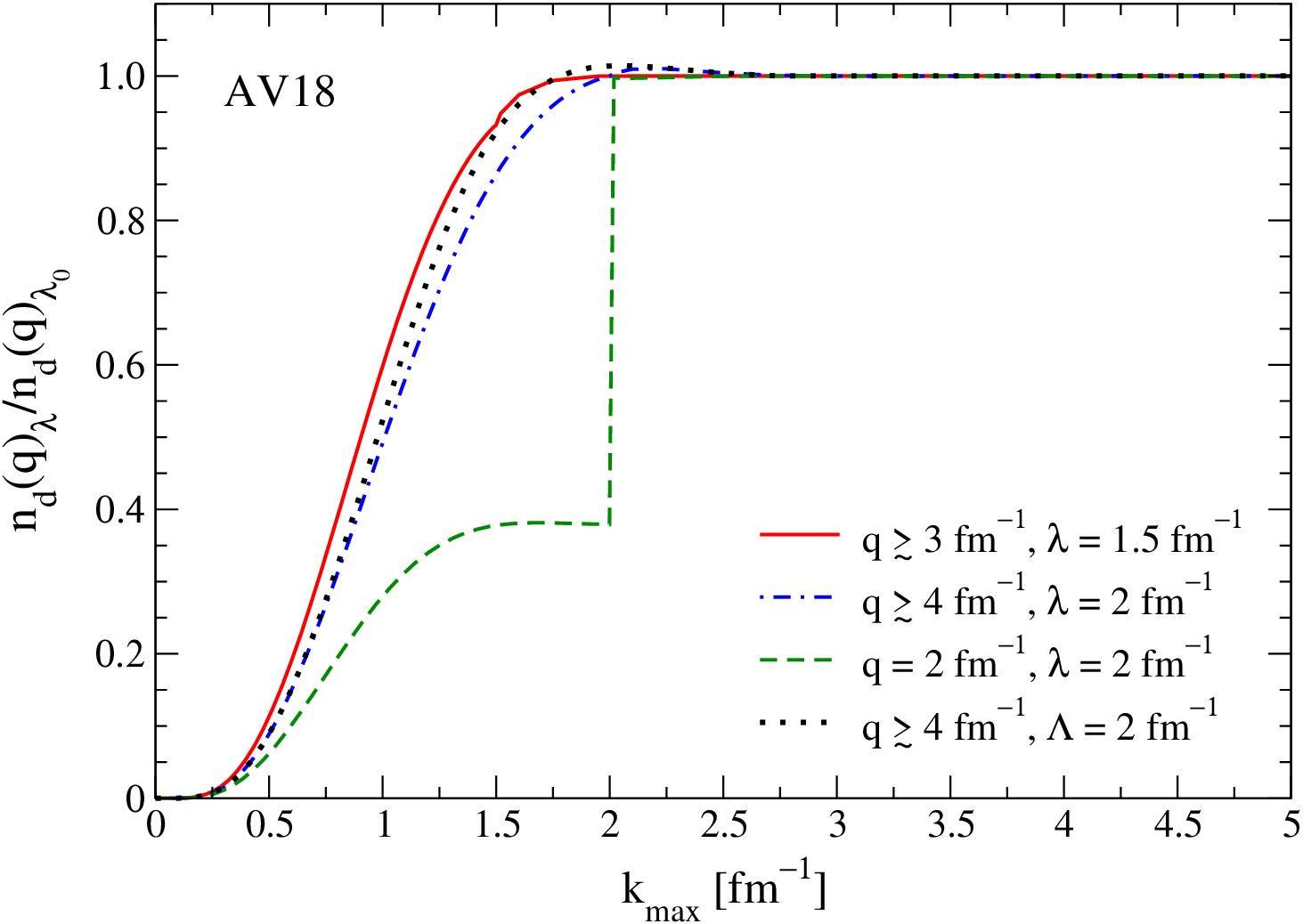}
\end{center}
\caption{Ratio of the deuteron momentum distribution at various momenta
$q$ evolved from the Argonne $v_{18}$ potential~\cite{AV18} via the SRG 
to the corresponding initial momentum 
distribution, as a function of the maximum momentum $\kmax$ in the deuteron 
wave functions in the numerator.  Note that the un-evolved Argonne 
$v_{18}$ result is simply a step function at q. For comparison, we also
show the result for a smooth-cutoff $\vlowk$ interaction with $\Lambda 
= 2\fmi$ and exponential regulator $n_{\rm exp}=2$.}
\label{fig:deutmd_accum}
\end{figure}

In Fig.~\ref{fig:deutmd}, we plot deuteron matrix elements
of $\adaggera$ for the Argonne 
$v_{18}$~\cite{AV18} and CD-Bonn~\cite{CDBonn} potentials, as well
as for two SRG and a smooth-cutoff $\vlowk$ interaction evolved from
Argonne $v_{18}$.
We emphasize again that matrix elements of this operator are not
measurable, so one should not ask which of these
is the ``correct'' momentum distribution in the deuteron; 
it is a potential- and scale-dependent quantity.
It is evident that the $\vsrg$ distributions have substantial
momentum components only for $k$ below $\lambda$, and that
the $\vlowk$ distribution is very similar to the corresponding
$\vsrg$ distribution.
Nevertheless, if we use the SRG- or RG-evolved operator
with these deuteron wave functions, we precisely reproduce the
momentum distribution from the original potential
at \emph{all momenta} and for \emph{all $\lambda$}.

This result by itself is guaranteed by construction.
The more interesting issue is where the strength of the matrix
element comes from.
For example, for the bare operator and the Argonne $v_{18}$ potential,
the momentum distribution at $q = 4\fmi$ comes entirely from
deuteron wave function components at that momentum.
But when $\lambda = 2\fmi$, it
is clear from Fig.~\ref{fig:deutmd}
that the deuteron does not have appreciable
momentum components
above $2.5\fmi$
(even though $\vsrg$ does near the diagonal \cite{Bogner:2006srg}).
In Fig.~\ref{fig:deutmd_accum}, we take the ratio of
the evolved operator evaluated with the evolved wave function at $q$
to the corresponding initial quantity, but include in the
numerator only momenta up to $\kmax$. The numerator is thus:
\beqn
\int_0^{\kmax}\! d{\bf k} \int_0^{\kmax}\! d{\bf k'}\,
\psi_d^\dagger({\bf k};\lambda) U_\lambda({\bf k,q}) 
U^\dagger_\lambda({\bf q,k'})
\psi_d({\bf k'};\lambda) \,.
\eeqn

We observe that for all $q$, the ratio approaches unity for large 
enough $\kmax$, as dictated by the unitary transformation. 
For $\lambda = 1.5\fmi$ or $2\fmi$, 
larger values of $q$
($q \gtrsim 3 \fmi$ or $q \gtrsim 4 \fmi$, respectively) 
give results approximately 
independent of $q$, with a smooth approach to unity by $\kmax \approx
1.3 \lambda$. This is consistent with the
operator $U_\lambda({\bf k,q})$ factorizing into $K_\lambda({\bf k}) \,
Q_\lambda({\bf q})$ for $k < \lambda$ and $q \gg \lambda$, and thus
the $q$ dependence cancels in the ratio. It remains to be seen whether
this factorization is a general feature
that can be understood using
operator product expansion ideas \cite{Lepage}. 
For small $q$, the original 
step function is essentially preserved (not shown). For $q$ of order $\lambda$,
there is a step behavior at $\lambda$ but some strength is shifted 
to lower $k,k'$. In all cases the flow of the operator 
strength weighted by $\psi_d$ 
is toward lower momenta. The ratio for the smooth-cutoff
$\vlowk$ is very similar at low $\kmax$ and for $\kmax$ approaching
$\Lambda$/$\lambda$. Finally, we have verified that taking matrix
elements 
using even a rough variational ansatz for the wave function also works 
fine, which serves as a check that there is no fine-tuning in the
evolved operator.

\section{Conclusions}
\label{sec:conclude}

Well-established renormalization theory tells us that low-energy
physics does not depend on the details of the high-energy dynamics.
The high-energy information we \emph{do}
need can be incorporated in nuclear interactions
using renormalization group methods designed to handle
similar problems in relativistic field theories and critical phenomena in
condensed matter systems.  This means that we can
decouple the physics of low energy from high energy,
drastically reducing the number of
explicit degrees of freedom required for precise non-perturbative
low-energy calculations.
	    
But while it is possible to decouple low- and high-energy physics,
it does not happen automatically. We have seen that when using
conventional NN potentials with strong short-range repulsion,
such as Argonne $v_{18}$~\cite{AV18}, there are momentum contributions
that are high on nuclear scales but must be included, or else even
very low-energy observables will be incorrect. We emphasize that the
need for high-momentum components in these particular calculations
\emph{does not} imply that the high-energy description is correct
(or measurable).

Two-nucleon interactions derived via the SRG explicitly decouple low-energy
physics from detailed high-momentum dynamics, allowing a clean
demonstration of these principles of renormalization.
Using SRG potentials, we have shown that
high-energy details in wave functions and operators are irrelevant
to low-energy observables. \emph{When we set the high-momentum components
to zero in $\vsrg$ potentials with low $\lambda$, there is no
noticeable
loss of information:} The low-energy phase shifts and expectation
values in the deuteron are practically unchanged. A corollary is 
that statements about
high-momentum parts in the deuteron or any other bound state depend on
the potential or scale. This
also holds for expectation values of other operators, such as the
one-pion exchange potential or three-nucleon interactions.
Finally, our results imply that there is also no problem with effective 
field theory interactions that predict zero phase shifts above a cutoff.

Using the SRG, operators can be evolved, including those associated
with high momenta in initial potentials, so that matrix elements in
low-energy states (like the deuteron) are unchanged.
We find for the momentum distribution that the operator strength flows
toward lower momentum, so the matrix elements do not change
 even after
contributions from high momenta are excluded. We also observe that
the ratio of evolved to initial deuteron momentum distribution is
nearly independent of momentum $q$ at fixed $\lambda$ for sufficiently
large $q$, which motivates
an investigation using an operator product expansion.

In the past, unitary transformations of NN potentials were applied
to calculations of few-body systems and nuclear matter.
Results for observables depended on the transformations and this was often
the context for discussing ``off-shell'' effects and which was the
``true potential''. From the modern perspective, this approach
is misleading at best. These transformations always lead to many-body
interactions, even if absent in the initial hamiltonian. While this
fact was clearly recognized in past investigations \cite{COESTER70},
many-body forces were usually neglected. 
The SRG is one promising
approach to evolve consistent three-nucleon interactions. 
Once the three-body SRG is implemented,
the tests of decoupling can be extended to systems with more than
two nucleons.

\begin{ack}
This work was supported in part by the National Science 
Foundation under Grant No.~PHY--0354916,  
the Department of Energy under Grant No. DE-FC02-07ER41457,
and the Natural Sciences and Engineering Research Council of Canada 
(NSERC). TRIUMF receives federal funding via a contribution agreement 
through the National Research Council of Canada.
\end{ack}

\end{document}